\documentclass[5p,times]{elsarticle}
\makeatletter
\def\ps@pprintTitle{%
 \let\@oddhead\@empty
 \let\@evenhead\@empty
 \def\@oddfoot{Published in Pattern Recognition, \url{https://doi.org/10.1016/j.patcog.2022.109064}\hfill}%
 \let\@evenfoot\@oddfoot}
\makeatother

\usepackage[utf8]{inputenc} 

\usepackage[table]{xcolor}
\usepackage{subcaption}
\usepackage{balance}
\usepackage{amsmath,amssymb}
\usepackage{url}

\usepackage{color}
\usepackage{xcolor}
\usepackage{bbm}
\usepackage{xspace}
\usepackage{graphicx}
\usepackage{longtable}
\usepackage{caption}
\usepackage{amsmath,amsthm}
\usepackage{multirow}
\usepackage{multicol}
\usepackage{booktabs}
\usepackage{url}
\usepackage{pifont}
\usepackage{color}
\usepackage{supertabular}
\usepackage[export]{adjustbox}
\usepackage{algorithm}
\usepackage[ruled,vlined,linesnumbered,algo2e]{algorithm2e}
\usepackage{algpseudocode}

\usepackage{amsmath}
\usepackage{amsthm,amsmath,amssymb}
\usepackage{mathrsfs}
\usepackage{bm}
\usepackage{subcaption}
\usepackage{hyperref}
\usepackage{CJKutf8}

\usepackage{cleveref}

\newcommand{\yes}{\ding{51}}%
\newcommand{\no}{\ding{55}}%
\usepackage{balance}

\newcommand{\vct}[1]{\ensuremath{\boldsymbol{#1}}}
\newcommand{\mat}[1]{\ensuremath{\mathbf{#1}}}

\newcommand{\edit}[1]{\textcolor{black}{#1}}

\newcommand{\myparagraph}[1]{\smallskip \noindent \textbf{#1}}
\newcommand{\ie}{{i.e.}\xspace}

\newcommand{\etal}{{et al.}\xspace}

\newcommand{\wrt}{{w.r.t.}\xspace}

\newcommand{\datasetname}{ImageNet-Patch\xspace}
\newcommand{\applypatch}{\ensuremath{\oplus}}
\newcommand{\numsamples}{J\xspace}
\newcommand{\sampleindex}{j\xspace}
\newcommand{\nummodels}{M\xspace}
\newcommand{\modelindex}{m\xspace}
\newcommand{\accuracyoperator}{\ensuremath{\mathcal{A}}}
\newcommand{\cleanacc}{C\xspace}
\newcommand{\robustacc}{R\xspace}
\newcommand{\successrate}{S\xspace}
\newcommand{\topk}{top-$k$\xspace}

\newcommand{\largescalenummodels}{127\xspace}
\newcommand{\numrobust}{3\xspace}
\newcommand{\numnonrobust}{3\xspace} 
\newcommand{\numgroups}{5\xspace}
\newcommand{\totmodels}{15\xspace} 
\newcommand{\whiteboxmodelsgroup}{\texttt{ENSEMBLE}\xspace}
\newcommand{\standardgroup}{\texttt{STANDARD}\xspace}
\newcommand{\robustgroup}{\texttt{ADV-ROBUST}\xspace}
\newcommand{\augmentationgroup}{\texttt{AUGMENTATION}\xspace}
\newcommand{\moredatagroup}{\texttt{MORE-DATA}\xspace}
\newcommand{\rot}{\texttt{rot}\xspace}
\newcommand{\loc}{\texttt{loc}\xspace}
\newcommand{\scl}{\texttt{scl}\xspace}
\newcommand{\various}{\texttt{various}\xspace}

\begin{document}

\begin{frontmatter}

\title{ImageNet-Patch: A Dataset for Benchmarking Machine Learning Robustness against Adversarial Patches}

\author[unica,pluribus]{Maura Pintor}

\author[unica]{Daniele Angioni}

\author[unica,pluribus]{Angelo Sotgiu}

\author[unica,pluribus]{Luca Demetrio}

\author[unica]{Ambra Demontis\corref{mycorrespondingauthor}}
\cortext[mycorrespondingauthor]{Corresponding author}
\ead{ambra.demontis@unica.it}

\author[unica,pluribus]{Battista Biggio}

\author[unige,pluribus]{Fabio Roli}

\address[unica]{University of Cagliari, Italy}
\address[unige]{University of Genova, Italy}
\address[pluribus]{Pluribus One, Italy}


\begin{abstract}

%
\edit{Adversarial patches are optimized contiguous pixel blocks in an input image that cause a  machine-learning model to misclassify it.
However, their optimization is computationally demanding, and requires careful hyperparameter tuning, potentially leading to suboptimal robustness evaluations.
To overcome these issues, we propose \datasetname, a dataset to benchmark machine-learning models against adversarial patches.
The dataset is built by first optimizing a set of adversarial patches against an ensemble of models, using a state-of-the-art attack that creates transferable patches. The corresponding patches are then randomly rotated and translated, and finally applied to the ImageNet data.
We use \datasetname to benchmark the robustness of \largescalenummodels models against patch attacks, and also validate the effectiveness of the given patches in the physical domain (i.e., by printing and applying them to real-world objects). We conclude by discussing how our dataset could be used as a benchmark for robustness, and how our methodology can be generalized to other domains. We open source our dataset and evaluation code at \url{https://github.com/pralab/ImageNet-Patch}.}

\end{abstract}







\end{frontmatter}

\section{Introduction}\label{sec1}

Understanding the security of machine-learning models is of paramount importance nowadays, as these algorithms are used in a large variety of settings, including security-related and mission-critical applications, to extract actionable knowledge from vast amounts of data. 
Nevertheless, such data-driven algorithms are not robust against adversarial perturbations of the input data~\cite{biggio13-ecml,szegedy14-iclr,carlini17-sp,madry18-iclr}.
In particular, attackers can hinder the performance of classification algorithms by means of \textit{adversarial patches}~\cite{brown2017adversarial}, i.e., contiguous chunks of pixels which can be applied to any input image to cause the target model to output an attacker-chosen class.
When embedded into input images, adversarial patches produce out-of-distribution samples.
The reason is that the injected patch induces a spurious correlation with the target label, which is likely to shift the input sample off the manifold of natural images.
Adversarial patches can be printed as stickers and physically placed on real objects, like stop signs that are then recognized as speed limits~\cite{eykholt2018robust}, and accessories that camouflage the identity of a person, hiding their real identity~\cite{sharif16-ccs,wei2022adversarial}. 
Therefore, the evaluation of the robustness against these attacks is of the uttermost importance, as they can critically impact real-world applications with physical consequences.

The only way to assess the robustness of a machine-learning system against adversarial patches is to generate and test them against the target model of choice. 
Adversarial patches are created by solving an optimization problem via gradient descent. However, this process is costly as it requires both querying the target model many times and computing the back-propagation algorithm until convergence is reached.
Hence, it is not possible to obtain a fast robustness evaluation against adversarial patches without avoiding all the computational costs required by their optimization process.
To further exacerbate the problem, adversarial patches should also be effective under different transformations, including translation, rotation and scale changes. 
This is required for patches to work also in the physical world, where it is impossible to place them in a controlled manner, i.e., to control the acquisition and environmental conditions.
Moreover, adversarial patches should also be transferable to different models, given that, in practice, the target model may not be exactly known to the attacker.

\edit{To overcome these issues, in this work we propose \datasetname, a dataset of pre-optimized adversarial patches that can be used to benchmark machine-learning models with small computational overhead.
This dataset is constructed on top of a subset of the validation set of the ImageNet dataset, coherently with other state-of-the-art benchmarks for robust models~\cite{croce2021robustbench}. It consists of $10$ patches that target $10$ different classes, applied on $5,000$ images each, for a total of $50,000$ samples.
We create these patches \edit{using the \textit{adversarial patch attack} proposed in~\cite{brown2017adversarial}}, which targets an ensemble of models to ensure that the resulting patches transfer well across different models (Section~\ref{sec:sec2_adv}).
The patches are also optimized to work under different rotation and translation. This makes them suited to stage physical attacks where the acquisition and environmental conditions cannot be controlled.}

\begin{figure*}[t]
    \centering
    \includegraphics[width=\textwidth]{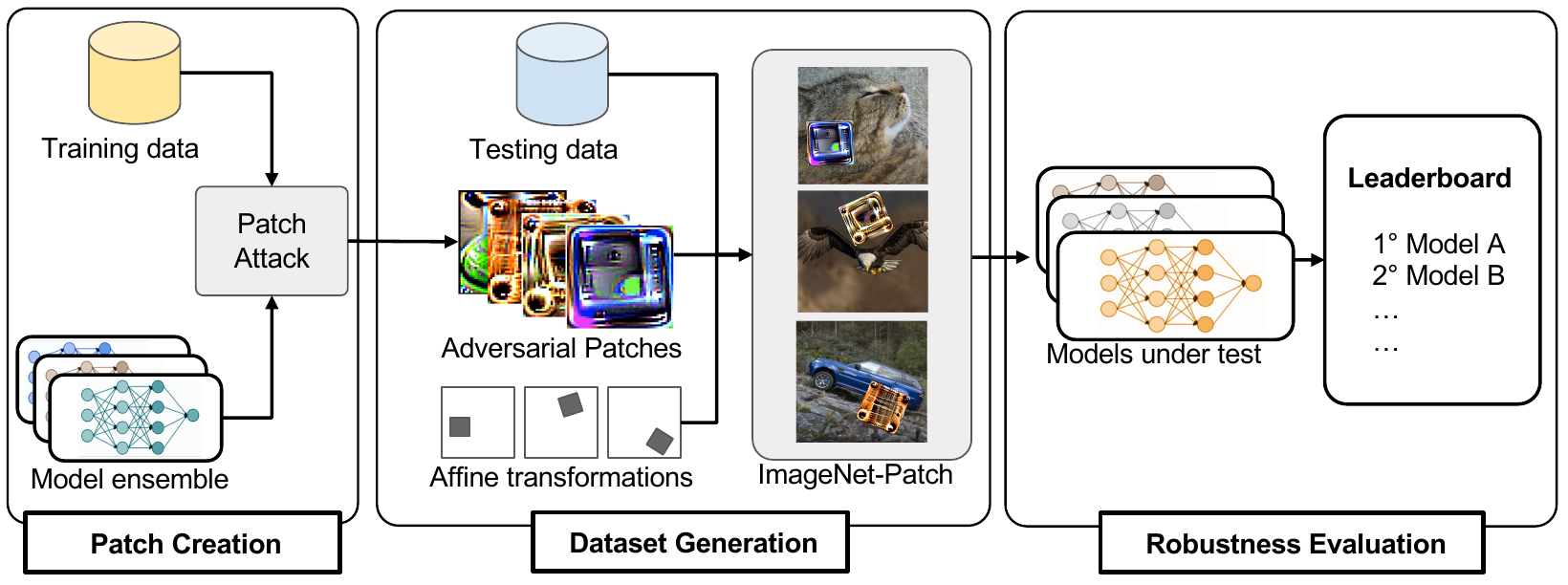}
    \caption{\edit{The three-step methodology followed to build our \datasetname benchmark.}}
    \label{fig:eval_robustness_dataset}
\end{figure*}

\edit{To build our benchmark, we follow a three-step methodology, as depicted in Figure~\ref{fig:eval_robustness_dataset}: (i) \textit{patch creation}, which amounts to optimizing adversarial (transferable) patches on the ImageNet dataset; (ii) \textit{dataset generation}, which consists of applying the aforementioned patches via random affine transformations; and (iii) \textit{robustness evaluation}, which amounts to assessing the robustness of the given models, and provides an appropriate ranking.
Even though the resulting robustness evaluation will be approximate, this process is extremely simple and fast, as newly-proposed defensive or robust learning mechanisms can be directly tested on the provided dataset, i.e., avoiding to repeat the patch-creation and dataset-generation steps (Section~\ref{sec:sec3_dataset}).}

We test the efficacy of ImageNet-Patch by evaluating \totmodels models that were not part of the initial ensemble as a test set, divided into \numnonrobust standard-trained models and \numrobust robustly-trained models, and we highlight the successful generalization of the patches to unseen models (Section~\ref{sec:sec4-exp}).
\edit{We also evaluate the effectiveness of the given patches in a real-world scenario by printing and applying them to three distinct physical objects, and acquiring 90 distinct images.
Our results demonstrate that this dataset can provide a quick yet approximate evaluation of the adversarial robustness of machine-learning models, avoiding the cumbersome task of re-optimizing the patches against each model.}
To foster reproducibility, we open-source the optimized patches along with the code used for evaluation.\footnote{\url{https://github.com/pralab/ImageNet-Patch}}

We conclude by discussing related work (Section~\ref{sec:sec5_related}), as well as the limitations and future directions of our work (Section~\ref{sec:sec6_conclusions}), envisioning a leaderboard of machine-learning models based on their robustness to ImageNet-Patch.

\section{Crafting Transferable Adversarial Patches} 
\label{sec:sec2_adv}



Attackers can compute adversarial patches by solving an optimization problem with gradient-descent algorithms~\cite{brown2017adversarial}.
Since these patches are meant to be printed and attached to real-world objects, their effectiveness should not be undermined by the application of affine transformations, like rotation, translation and scale, that are unavoidable when dealing with this scenario.
For example, an adversarial patch placed on a traffic sign should be invariant to scale changes to remain effective while an autonomous driving car approaches the traffic sign, or to camera rotation when taking pictures.
Hence, the optimization process must include these perturbations as well, to force such invariance inside the resulting patches.
Also, adversarial patches can either generate a general misclassification, namely an \emph{untargeted} attack, or force the model to predict a specific class, namely a \emph{targeted} attack.
In this paper, we focus on the latter, and we consider a patch effective if it is able to correctly pilot the decision-making of a model toward an intended class.

More formally, targeted adversarial patches are computed by solving the following optimization problem:
\begin{equation}
    \min_{\vct \delta} \mathbb{E}_{\mat A \sim \mathcal{T}}\left[\sum_{\sampleindex=1}^{\numsamples}\mathcal{L}(\vct x_\sampleindex \applypatch \mat A \vct \delta, y_t; \vct \theta)\right],
    \label{eq:objfunctionsingle}
\end{equation}
where $\vct \delta$ is the adversarial patch to be computed, $\vct x_\sampleindex$ is one of $\numsamples$ samples of the training data, $y_t$ is the target label,\footnote{The same formulation holds for crafting untargeted attacks, by simply substituting the target label $y_t$ with the ground truth label of the samples $y$, and inverting the sign of the loss function.} $\vct \theta$ is the targeted model, $\mat A$ is an affine transformation randomly sampled from a set of affine transformations $\mathcal{T}$, $\mathcal{L}$ is a loss function of choice, that quantifies the classification error between the target label and the predicted one and $\applypatch$ is a function that applies the patch on the input images.
The latter is defined as: $\vct x \applypatch \vct \delta = (\vct 1 - \vct \mu) \odot \vct x + \vct \mu \odot \vct \delta$, where we introduce a mask $\vct \mu$ that is a tensor with the same size of the input data $\vct x$, and whose components are ones where the patch should be applied and zeros elsewhere~\cite{karmon2018lavan}.
This operator is still differentiable, as it is constructed by summing differentiable functions themselves; thus, it is straightforward to obtain the gradient of the loss function with respect to the patch.

To produce a dataset that can be used as a benchmark for an initial robustness assessment, with adversarial patches effective regardless of the target model, we \edit{leverage the technique proposed by Brown et al.~\cite{brown2017adversarial}}, that considers an ensemble of differentiable models inside the optimization process.
This addition forces the optimization algorithm to find effective solutions against all the ensemble models, boosting the transferability of the produced adversarial patches. 
Namely, the ability of the adversarial patch optimized against a model (or a set of them) to be effective against different models.
Hence, the loss function to be minimized can be written as:
\begin{equation}
    \min_{\vct \delta} \mathbb{E}_{\mat A \sim \mathcal{T}}\left[\sum_{\modelindex=1}^{\nummodels}\sum_{\sampleindex=1}^{\numsamples}\mathcal{L}(\vct x_\sampleindex \applypatch \mat A \vct \delta, y_t; \vct \theta_\modelindex)\right],
    \label{eq:objfunctionensemble}
\end{equation}
where we modified the original formulation in Equation~\ref{eq:objfunctionsingle} to minimize the loss $\mathcal{L}$ over a set of $M$ models, respectively parameterized via $\vct \theta_1, ..., \vct \theta_\nummodels$.
\begin{figure*}[t]
    \centering
    \includegraphics[width=\linewidth]{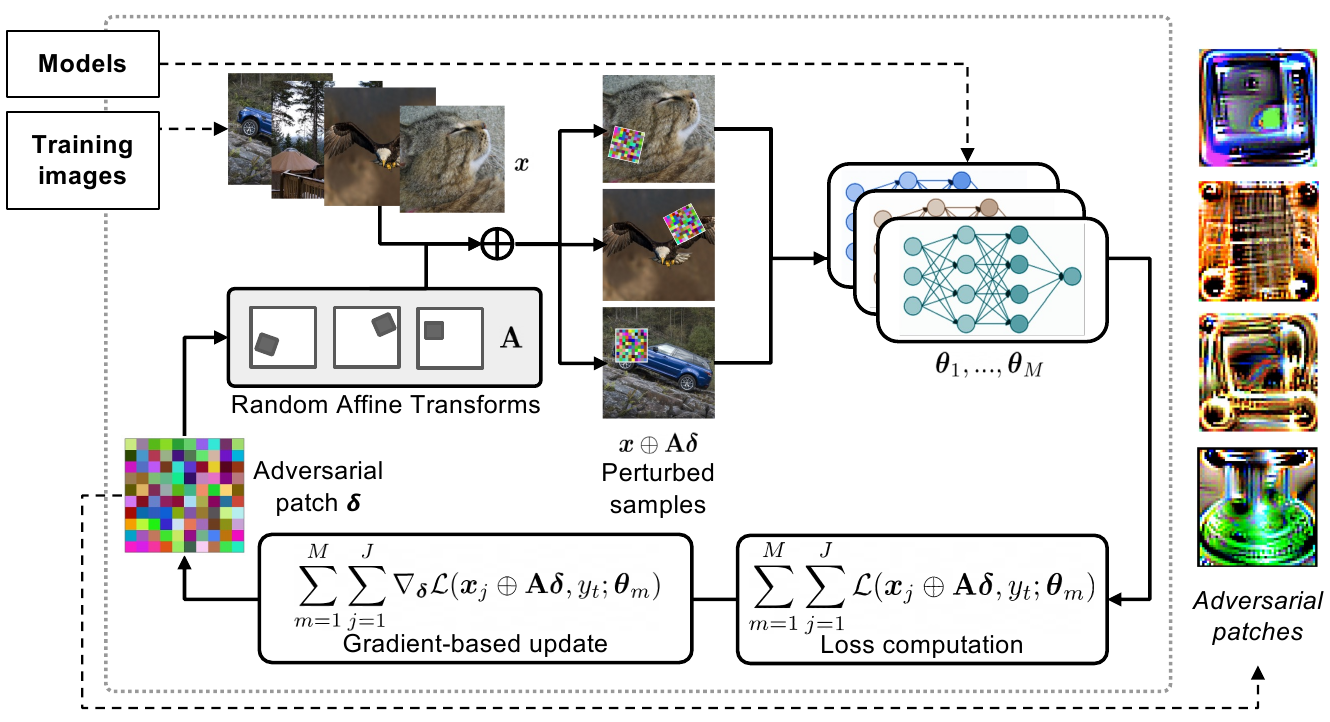}
    \caption{The optimization process, graphically described. At each step, we apply the patch to be optimized with random affine transformations on sample images, and we compute the scores of the ensemble. Hence, the algorithm computes the update step through gradient descent on the loss function \wrt the patch.}
    \label{fig:optimization_schema}
\end{figure*}


The objective function defined in Equation~\ref{eq:objfunctionensemble} can be optimized through gradient-descent techniques, and thus we use Algorithm~\ref{algo:advPatch} for minimizing it.
After having randomly initialized the patch (line \ref{line:init_patch}), we loop through the number of intended epochs (line~\ref{line:epochs}), and the samples of the training data (line~\ref{line:samples}).
In each epoch, we sample a random affine transformation that will be applied to the patch (line~\ref{line:init_affine}).
\edit{We differ from the original formulation of Brown et. al~\cite{brown2017adversarial}, as we solely consider rotations and translations.}
We iterate over all models of the ensemble (line~\ref{line:n_models}) to calculate the loss by accumulating its gradient \wrt the patch (line~\ref{line:accumulate}), and using it to update the patch at the end of each epoch (line~\ref{line:optimize}).
After all the epochs have been consumed, the final adversarial patch is returned (line~\ref{line:return}).
If the number of training samples is large, this algorithm can be easily generalized to a more efficient version using the gradient computed on a mini-batch to perform the updates, \ie repeating the steps 3-8 for each batch of the training data.
We present a graphical representation of our procedure in Figure~\ref{fig:optimization_schema}.

\begin{algorithm}[t]
    \SetKwInOut{Input}{Input}
    \SetKwInOut{Output}{Output}
    \SetKwComment{Comment}{$\triangleright$\ }{}
    \DontPrintSemicolon
    \Input{$\vct x$, the training dataset containing $\numsamples$ images; $y_t$, the target class; $\vct \theta_1,.., \vct \theta_\nummodels$, the ensemble of models; $\gamma$, the learning rate; $N$, the number of epochs.}
    \Output{$\vct \delta$, the adversarial patch}
    $\vct \delta \sim U(0,1)$ \Comment*[r]{Initialize patch with uniform distribution}\label{line:init_patch}
    \For{$i \in [1, N]$}{\label{line:epochs}
        $\vct g \leftarrow 0$ \Comment*[r]{Initialize gradient update for epoch $i$}\label{line:init_grad}
        \For{$\sampleindex \in [1, \numsamples]$}{\label{line:samples}
            $\mat A \leftarrow \texttt{random-affine}()$ \Comment*[r]{Initialize transformation}\label{line:init_affine}
            \For{$\modelindex \in [1, \nummodels]$}{\label{line:n_models}
                $\vct g \leftarrow \vct g + \frac{1}{\nummodels \numsamples} \nabla_{\vct \delta} \mathcal{L}(\vct x_\sampleindex \applypatch \mat A \vct \delta, y_t; \vct \theta_\modelindex)$ \Comment*[r]{Accumulate gradients}\label{line:accumulate}
            }
        }
        
        $\vct \delta \leftarrow \vct \delta - \gamma \vct g$ \Comment*[r]{Optimize patch}\label{line:optimize}
        
    }
    \textbf{return} $\vct \delta$ \Comment*[r]{Return optimized patch}\label{line:return}
\caption{Optimization of adversarial patches on an ensemble of models}
\label{algo:advPatch}
\end{algorithm}

\section{The ImageNet-Patch Dataset}
\label{sec:sec3_dataset}
We now illustrate how we apply our methodology to generate the \datasetname dataset that will be used to evaluate the robustness of classification models against patch attacks.

\myparagraph{The Baseline Dataset.}
We start from the validation set of the original ImageNet database,\footnote{\url{https://www.image-net.org/challenges/LSVRC/index.php}} containing $1,281,167$ training images, $50,000$ validation images and $100,000$ test images, divided into $1,000$ object classes. 
From the validation set, we select a test set of $5,000$ images that matches exactly the ones used in RobustBench~\cite{croce2021robustbench} for testing model robustness against adversarial attacks. This allows us not only to provide a direct comparison with the RobustBench framework, but also to easily add our benchmark to it. 
We create the corpus of images used to optimize adversarial patches from the remaining part of the ImageNet validation set, excluding the images used for the test set, and randomly sampling $20$ images from different classes.
Each patch is then optimized on these samples except the images of the target class of the attack.
To clarify, if the attack is targeting the class ``cup'', we select one image for each of 20 different classes selected from the remaining 999 classes of the ImageNet dataset.

\myparagraph{The ImageNet-Patch Dataset.} 
We now \edit{discuss how we generate the \datasetname dataset. 
We apply the methodology proposed by Brown et al.~\cite{brown2017adversarial} that} optimizes adversarial patches on an ensemble of chosen models, and we select three deep neural network architectures trained on the ImageNet dataset, namely AlexNet~\cite{krizhevsky2012imagenet}, ResNet18~\cite{he2016deep} and SqueezeNet~\cite{iandola2016squeezenet}.
We leverage the pretrained models available inside the PyTorch TorchVision zoo,\footnote{\url{https://pytorch.org/vision/master/models.html}} that are trained to take in input RGB images of size $224 \times 224$.


We run Algorithm~\ref{algo:advPatch} to create squared patches with a size of $50 \times 50$ pixels, with a learning rate of $1$, $20$ training samples selected as previously described, $5000$ training epochs, and using the cross-entropy as the loss function of choice.
We consider rotation and translation as the applied affine transformations during the optimization of the patch, constraining rotations up to $\pm \frac{\pi}{8}$ to mimic the setup applied by Brown et al.\cite{brown2017adversarial}, and translations to a shift of $\pm 68$ pixels on both axes from the center of the image.
The latter is a heuristic constraint, as we want to avoid corner cases where the adversarial patch is too close to the boundaries of the image.
We also keep the size of the adversarial patch fix to $50 \times 50$ pixels during the optimization process.

\begin{figure*}[t]
    \centering
        \includegraphics[width=0.99\linewidth]{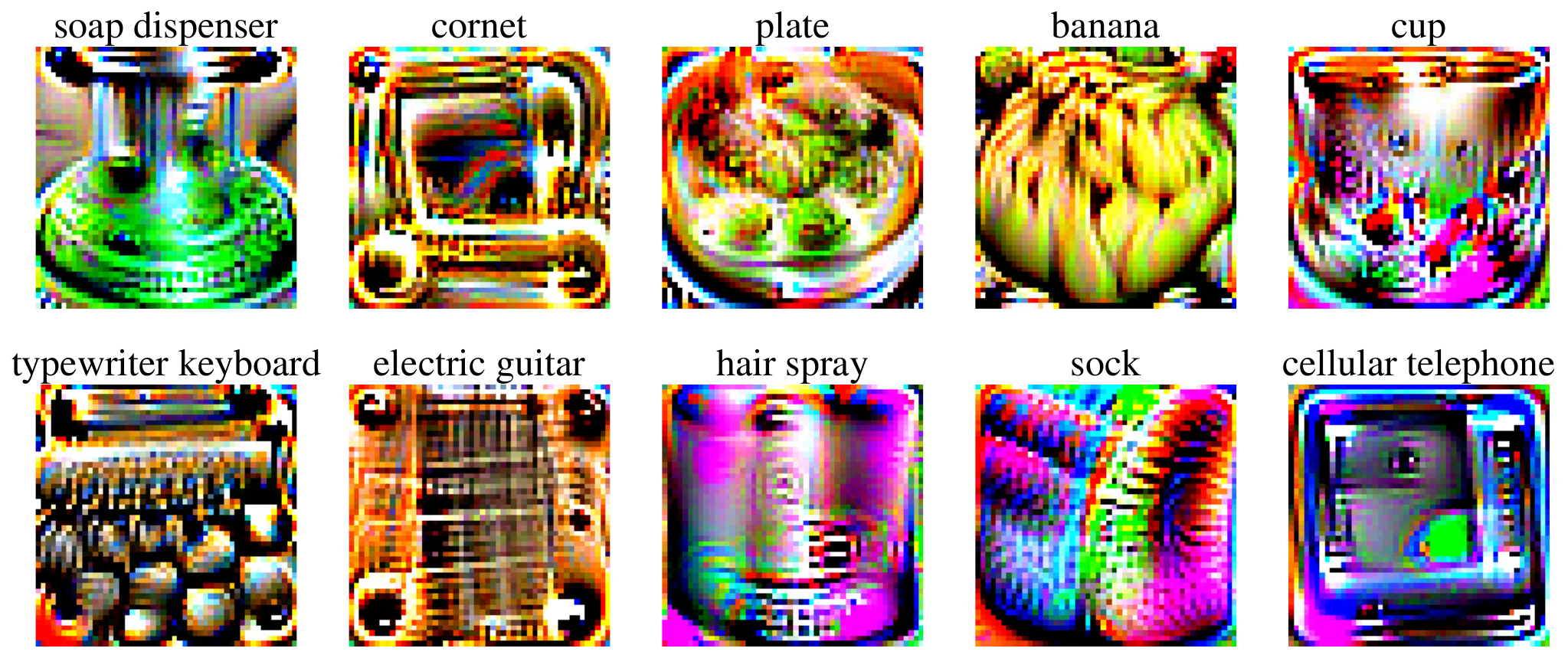}
    \caption{The 10 optimized adversarial patches, along with their target labels.}
    \label{fig:imagenet_patches}
\end{figure*}

\begin{figure*}[t]
    \centering
        \includegraphics[width=\textwidth]{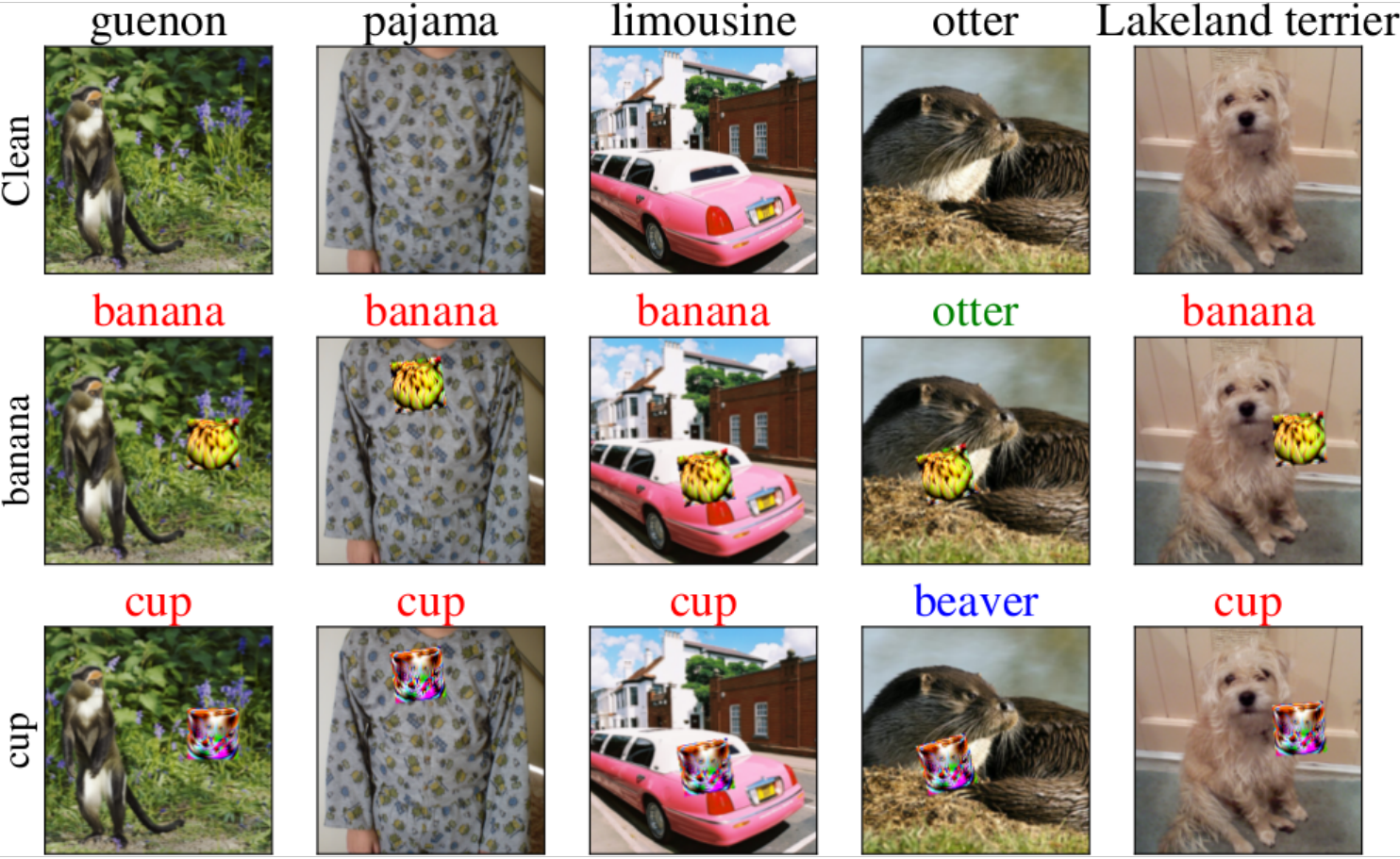}
    \caption{A batch of clean images initially predicted correctly by a SqueezeNet~\cite{iandola2016squeezenet} model, and its perturbation with 2 different adversarial patches. Each row contains the original image with a different patch, whose target is displayed in the left. The predictions are shown on top of each of the samples, in \textit{\textcolor{green}{green}} for correct prediction, \textit{\textcolor{blue}{blue}} for misclassification, and in \textit{\textcolor{red}{red}} for a prediction that ends up in the target class of the attack.}
    \label{fig:imagenet_images}
\end{figure*}

We optimize 10 different patches with these settings, targeting 10 different classes of the ImageNet dataset (``soap dispenser'', ``cornet'', ``plate'', ``banana'', ``cup'',  ``typewriter keyboard'', ``electric guitar'', ``hair spray'',  ``sock'', ``cellular phone''). The resulting patches are shown in Figure~\ref{fig:imagenet_patches}.
We apply such patches to each of the $5,000$ images in the test set along with random affine transformations, generating a dataset of $50,000$ perturbed images with adversarial patches.
We depict some examples in Figure~\ref{fig:imagenet_images}.


\section{Experimental Analysis}\label{sec:sec4-exp}

We now showcase experimental results related to the robustness evaluation through the usage of our \datasetname dataset.
We first explain the metrics (Section \ref{subsec:eval_metrics}), and which models we consider for evaluating our dataset (Section \ref{subsec:eval_protocol}).
We then proceed in detailing the results of our experiments (Section \ref{subsec:exp_results}), by considering the previously introduced metrics, and lastly we show the same measurements but extended to a large-scale model selection (Section~\ref{subsec:large_scale}).

\subsection{Evaluation Metrics}
\label{subsec:eval_metrics}
We evaluate the evasion performance of the \datasetname dataset by considering three metrics: (i) the \textit{clean accuracy}, which is the accuracy of the target model in absence of attacks; (ii) the \textit{robust accuracy}, which is the accuracy of the target model in presence of adversarial patches; and (iii) the \textit{success rate} of a patch, that measures the percentage of samples for which the patch successfully altered the prediction of the target model toward the intended class.

\myparagraph{Clean Accuracy.} We denote with the operator $\accuracyoperator_k(\vct x, y; \vct \theta)$ the \topk accuracy, \ie by inspecting if the desired class $y$ appears in the set of $k$ highest outputs of the classification model $\vct \theta$ when receiving the sample $\vct$ x as input. 
We then use this operator for defining the clean accuracy $\cleanacc_{k}$, as
\(
\cleanacc_{k} = \underset{
(\vct x, y) \sim \mathcal{D}_{test}}{\mathbb{E}}\left[\accuracyoperator_k(\vct x, y ;\vct \theta)\right],
\)
and the other metrics that we use for our experimental evaluation.

\myparagraph{Robust Accuracy.} We define the value $\robustacc_k$ as the \topk accuracy on the images after the application of the patch with the random rototranslation transformations, formalized as
\(
\robustacc_k = \underset{
\begin{subarray}{c}
(\vct x, y) \sim \mathcal{D}_{test}\\ 
\mat A \sim \mathcal{T}\end{subarray}
}{\mathbb{E}}\left[\accuracyoperator_k(\vct x \applypatch \mat A \vct \delta, y ;\vct \theta)\right]
\).

\myparagraph{Success Rate.} We define the value $\successrate_k$ as the success rate of the attack, \ie the \topk accuracy on the target label $y_t$ instead of the ground truth label $y$, formalized as \(
\successrate_k = \underset{
\begin{subarray}{c}
(\vct x, y) \sim \mathcal{D}_{test}\\ 
\mat A \sim \mathcal{T}\end{subarray}
}{\mathbb{E}}\left[\accuracyoperator_k(\vct x \applypatch \mat A \vct \delta, y_t ;\vct \theta)\right]
\)
We evaluate these three metrics for $k = 1,5,10$.

\subsection{Evaluation Protocol}
\label{subsec:eval_protocol}

To evaluate the effectiveness of the patches, we test our \datasetname dataset against \largescalenummodels deep neural networks trained on the ImageNet dataset.
To facilitate the discussion, we group the models in \numgroups groups, namely the \whiteboxmodelsgroup, \standardgroup, \robustgroup, \augmentationgroup, \moredatagroup groups. 
In a first analysis, we consider \totmodels models to discuss results in detail, and further extend the analysis with a large-scale analysis, presented in Section~\ref{subsec:large_scale}.
In particular, we consider the three models used for the ensemble, AlexNet~\cite{krizhevsky2012imagenet}, ResNet18~\cite{he2016deep} and SqueezeNet~\cite{iandola2016squeezenet}, as the first group, \whiteboxmodelsgroup.
We consider for the second group, \standardgroup,
\numnonrobust standard-trained models, that are  GoogLeNet~\cite{Szegedy2015GoingDW}, MobileNet~\cite{Howard2019SearchingFM} and Inception v3~\cite{Szegedy2016RethinkingTI}, available in PyTorch Torchvision. 
We then consider \numrobust robust-trained models as the \robustgroup available on RobustBench, specifically a ResNet-50 proposed by Salman et al.~\cite{DBLP:conf/nips/SalmanIEKM20}, a ResNet-50 proposed by Engstrom et al.~\cite{robustness} and a ResNet-50 proposed by Wong et al.~\cite{Wong2020Fast}.
We also additionally consider a set of 6 models from the ImageNet Testbed repository\footnote{\url{https://github.com/modestyachts/imagenet-testbed}} proposed by Taori \etal~\cite{taori2020measuring}, to analyze the effects of non-adversarial augmentation techniques and of training on bigger datasets. 
We select 3 models specifically trained for being robust to common image perturbations and corruptions, namely the models proposed by Zhang et al.~\cite{zhang2019shiftinvar}, Hendrycks et al~\cite{hendrycks2021many}, and Engstrom et al~\cite{engstrom2019exploring}, that we group as \augmentationgroup group. 
We further select other 3 models, namely two of the ones proposed by Yalniz et al.~\cite{yalniz2019billionscale} and one proposed by Mahajan et al.~\cite{wslimageseccv2018}, that have been trained on datasets that utilize substantially more training data than the standard ImageNet training set. We group these last models as the \moredatagroup group.
Lastly, the \standardgroup, \robustgroup, \augmentationgroup, and \moredatagroup groups will be referred as the \emph{Unknown} models, since they are not used while optimizing the adversarial patches.

\subsection{Experimental Results}
\label{subsec:exp_results}
\begin{figure*}[t]
    \centering
        \includegraphics[width=0.99\textwidth]{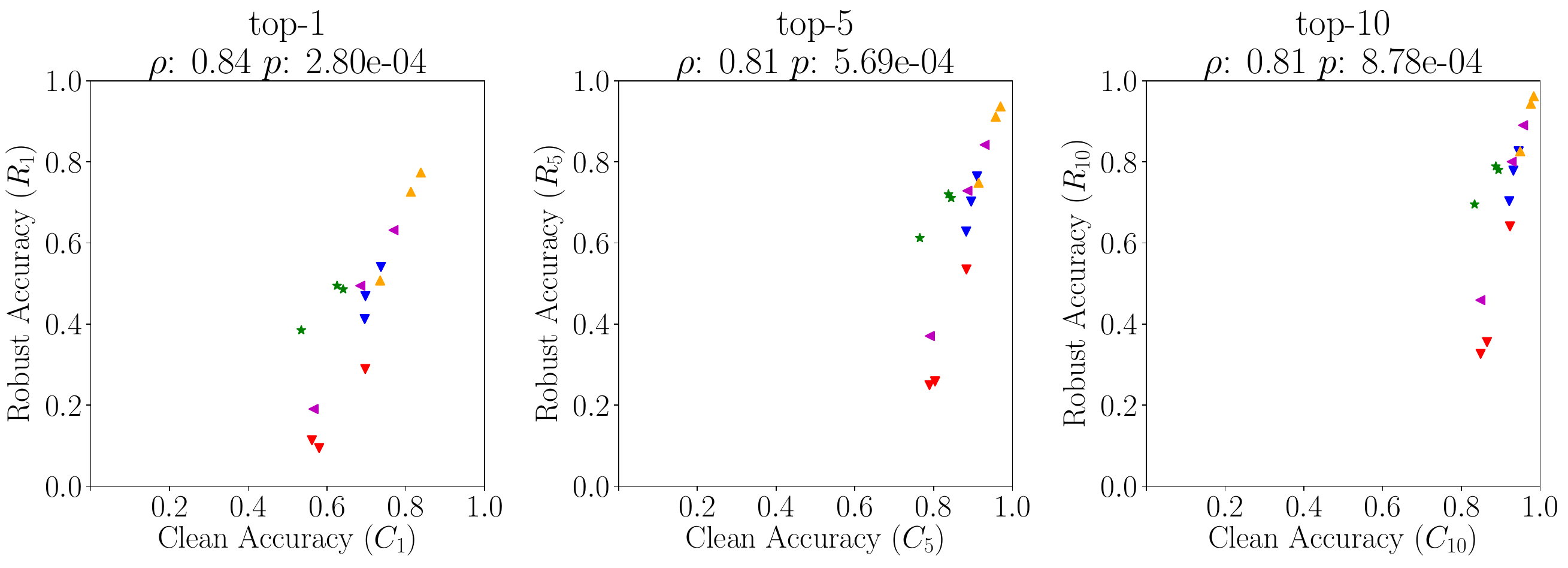}
        \includegraphics[width=0.99\textwidth]{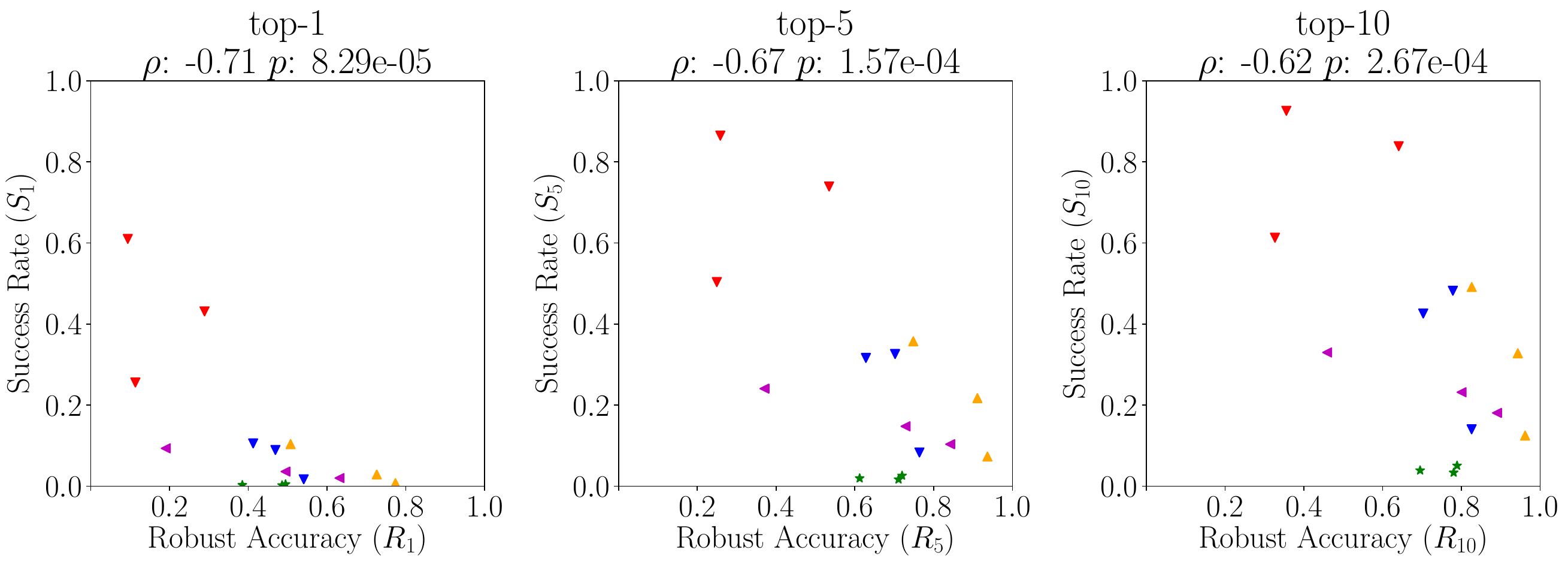}
        \includegraphics[width=0.99\textwidth]{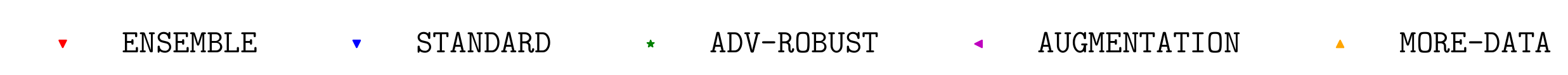}
    \caption{Analysis for results shown in Table~\ref{tab:patch-attacks}. \textit{Top row}: top-1 (left), top-5 (center), and top-10 (right) clean accuracy vs robust accuracy. \textit{Bottom row}: top-1 (left), top-5 (center), and top-10 (right) robust accuracy vs attack success rate. The Pearson correlation coefficient $\rho$ and the $p$-value are also reported for each plot.}
    \label{fig:subset_results}
\end{figure*}

We now detail the effectiveness of our dataset against the groups we have isolated, according to the chosen metrics. The results are reported in Table~\ref{tab:results_groups} and Figure~\ref{fig:subset_results}, where we confront the relation between clean and robust accuracy, and also between robust accuracy and success rate.

\myparagraph{Evaluation of Known Models.} The \whiteboxmodelsgroup group of models is characterized by low robust accuracy and the highest success rate of the adversarial patch, as expected, given that we optimize our adversarial patches to specifically mislead these models (they are part of the training ensemble).

\myparagraph{Evaluation of Unknown Models.}
These models are not part of the ensemble used to optimize the adversarial patches.
First of all, all of them highlight a good clean accuracy on our clean test set of images.

The \standardgroup group is characterized by a modest decrement of the robust accuracy, highlighting errors caused by the patches.
The success rate is lower compared to those exhibited by the \whiteboxmodelsgroup group, since patches are not optimized on these models, but it raises considerably when considering different top-k results.
This means that, even if the target class is not the predicted one, its confidence is still significantly increased.

\edit{The \robustgroup group is characterized by a drop of robust accuracy similar to the \standardgroup group, but with an almost-zero success rate for the adversarial patches.
This means that the predictions of robust models are still wrong, but they do not coincide with the target class.}

The \augmentationgroup group contains mixed results, shifting from a modest to a severe drop in terms of robust accuracy, associated with an increment of the success rate, which is slightly less than that achieved by the \standardgroup group.
\edit{This might imply that data augmentation helps the model to improve clean accuracy, but performance drops when dealing with adversarial noise.}

Lastly, the \moredatagroup group scores the best in terms of both clean and robust accuracy while the success rate of the adversarial patches is similar to the \augmentationgroup group results.

\begin{table*}[t]
\renewcommand{\arraystretch}{1.5}
\resizebox{0.99\linewidth}{!}{%
\begin{tabular}{l|l|lll|lll|lll|}

\cline{3-11}
\multicolumn{2}{c}{} &  \multicolumn{3}{|c}{\textbf{top-1}} & \multicolumn{3}{c}{\textbf{top-5}} & \multicolumn{3}{c|}{\textbf{top-10}} \\
\cline{2-11}
   &                     \textbf{Model} &            $\cleanacc_{1}$ & $\robustacc_{1}$ & $\successrate_{1}$ & $\cleanacc_{5}$ & $\robustacc_{5}$ & $\successrate_{5}$ & $\cleanacc_{10}$ & $\robustacc_{10}$ & $\successrate_{10}$ \\
\hline \hline
\multirow{3}{*}{\rotatebox[origin=c]{90}{\whiteboxmodelsgroup}}
&                      AlexNet~\cite{krizhevsky2012imagenet} &           0.562 &     0.113 &     0.256 &               0.789 &     0.250 &     0.504 &                0.849 &      0.327 &      0.613 \\
&                     ResNet18~\cite{he2016deep} &           0.697 &     0.289 &     0.431 &               0.883 &     0.535 &     0.739 &                0.923 &      0.641 &      0.839 \\
&                   SqueezeNet~\cite{iandola2016squeezenet} &           0.580 &     0.094 &     0.610 &               0.804 &     0.259 &     0.865 &                0.865 &      0.355 &      0.926 \\
\cline{1-11}
\multirow{3}{*}{\rotatebox[origin=c]{90}{\standardgroup}}
&                    GoogLeNet~\cite{Szegedy2015GoingDW} &           0.697 &     0.469 &     0.090 &               0.895 &     0.702 &     0.326 &                0.932 &      0.778 &      0.482 \\
&                    MobileNet~\cite{Howard2019SearchingFM} &           0.737 &     0.541 &     0.017 &               0.910 &     0.764 &     0.083 &                0.945 &      0.826 &      0.141 \\
&                    Inception v3~\cite{Szegedy2016RethinkingTI} &           0.696 &     0.412 &     0.106 &               0.883 &     0.628 &     0.317 &                0.921 &      0.703 &      0.426 \\
\cline{1-11}
\multirow{3}{*}{\rotatebox[origin=c]{90}{\robustgroup}}
&       Engstrom et al.~\cite{robustness} &           0.625 &     0.495 &     0.005 &               0.838 &     0.720 &     0.026 &                0.887 &      0.789 &      0.051 \\
&            Salman et al.~\cite{DBLP:conf/nips/SalmanIEKM20} &           0.641 &     0.486 &     0.003 &               0.845 &     0.711 &     0.017 &                0.894 &      0.780 &      0.034 \\
&                 Wong et al.~\cite{Wong2020Fast} &           0.535 &     0.385 &     0.003 &               0.765 &     0.612 &     0.020 &                0.833 &      0.695 &      0.039 \\
\cline{1-11}
\multirow{3}{*}{\rotatebox[origin=c]{90}{\texttt{AUGM.}}}
&                Zhang et al.~\cite{zhang2019shiftinvar} &           0.566 &     0.191 &     0.093 &               0.790 &     0.370 &     0.241 &                0.848 &      0.459 &      0.330 \\
&        Hendrycks et al~\cite{hendrycks2021many} &           0.769 &     0.632 &     0.020 &               0.929 &     0.842 &     0.104 &                0.956 &      0.890 &      0.181 \\
& Engstrom et al~\cite{engstrom2019exploring} &           0.684 &     0.495 &     0.036 &               0.886 &     0.729 &     0.148 &                0.928 &      0.800 &      0.232 \\
\cline{1-11}
\multirow{3}{*}{\rotatebox[origin=c]{90}{\moredatagroup}}
&      Yalniz et al.~\cite{yalniz2019billionscale}-a &          0.813 &     0.726 &     0.029 &               0.958 &     0.911 &     0.217 &                0.976 &      0.943 &      0.328 \\
& Yalniz et al.~\cite{yalniz2019billionscale}-b &          0.838 &     0.774 &     0.008 &               0.970 &     0.936 &     0.073 &                0.984 &      0.962 &      0.125 \\
&               Mahajan et al.~\cite{wslimageseccv2018} &          0.735 &     0.507 &     0.104 &               0.914 &     0.748 &     0.357 &                0.949 &      0.826 &      0.491 \\
\bottomrule

\end{tabular}
}
\caption{Evaluation of the \datasetname dataset using the chosen metrics, as described in Section~\ref{subsec:eval_protocol}.
On the rows, we list the 15 models used for testing, divided into the isolated groups.
On the columns, we detail the clean accuracy, the robust accuracy and the success rate of the adversarial patch, repeated for top-1,5, and 10 accuracy.}
\label{tab:results_groups}
\end{table*}

\subsection{Large-scale Analysis}
\label{subsec:large_scale}
\begin{figure*}[t]
    \centering
        \includegraphics[width=0.99\textwidth]{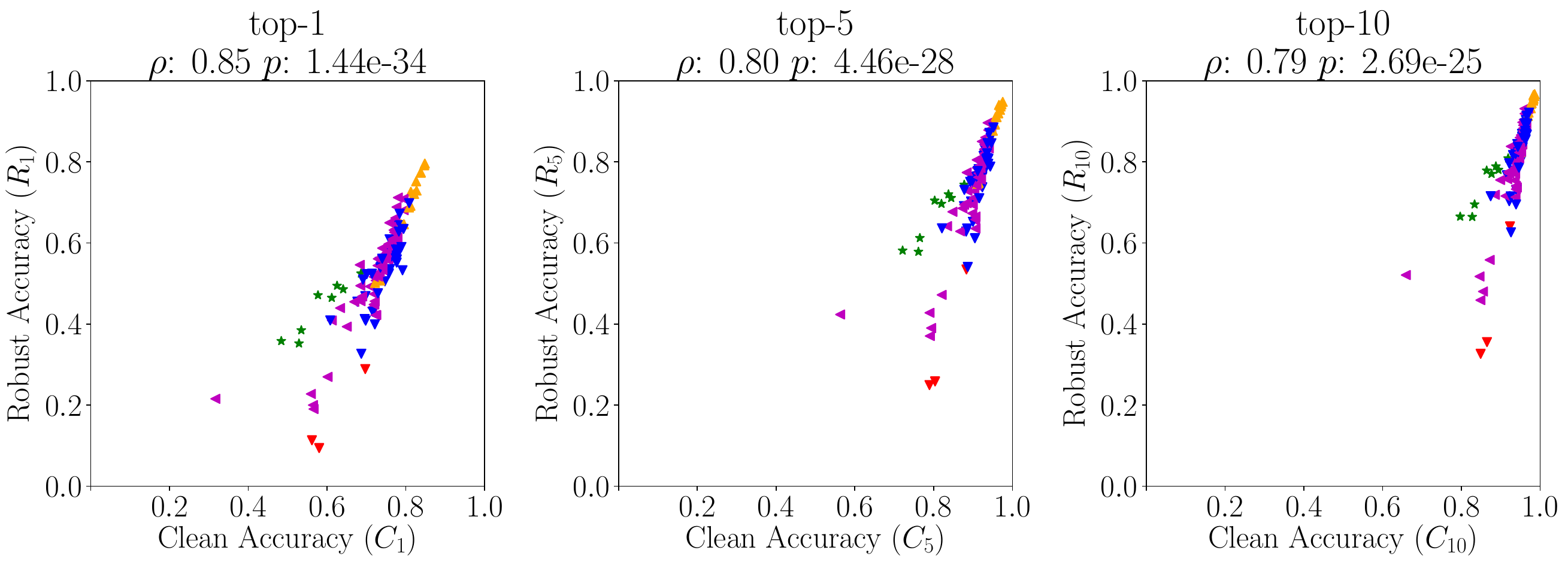}
        \includegraphics[width=0.99\textwidth]{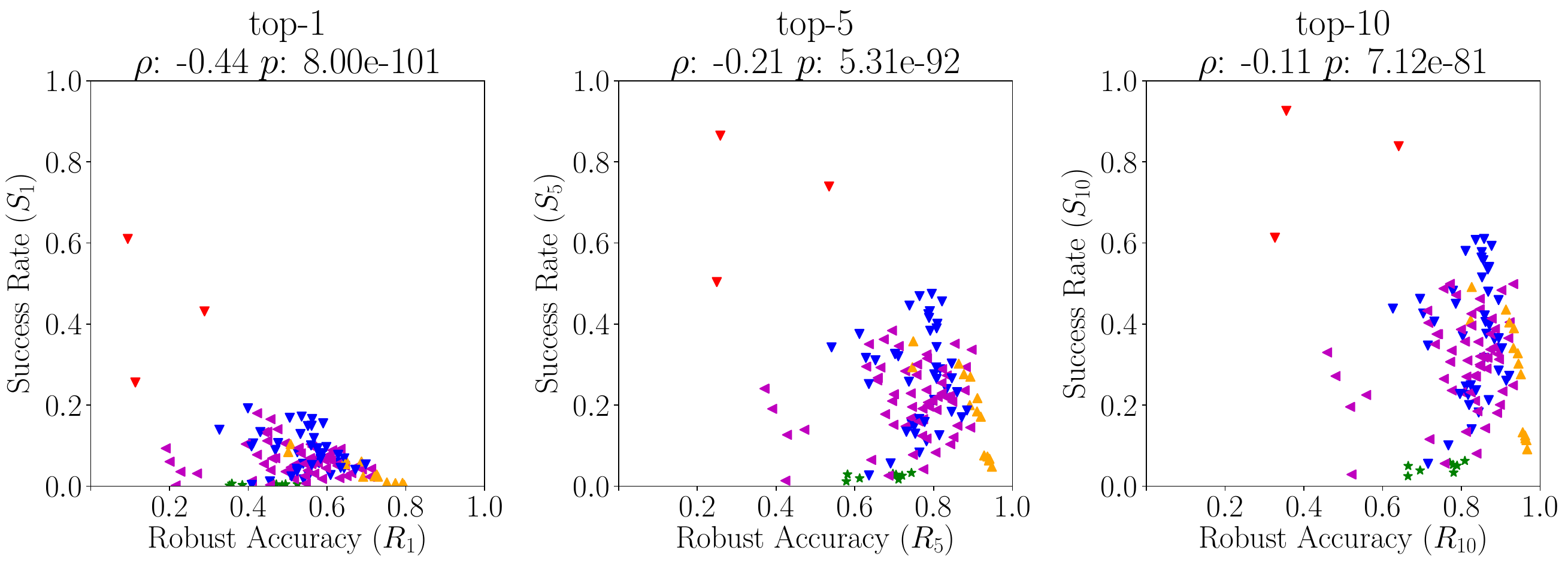}
        \includegraphics[width=0.99\textwidth]{figures/experiments/legend.pdf}
    \caption{Results of our large-scale analysis on \largescalenummodels publicly-released models. \textit{Top row}: top-1 (left), top-5 (center), and top-10 (right) clean accuracy vs robust accuracy. \textit{Bottom row}: top-1 (left), top-5 (center), and top-10 (right) robust accuracy vs attack success rate. The Pearson correlation coefficient $\rho$ and the $p$-value are also reported for each plot.}
    \label{fig:large-scale}
\end{figure*}

\begin{figure*}[t]
    \centering
        \includegraphics[width=0.99\textwidth]{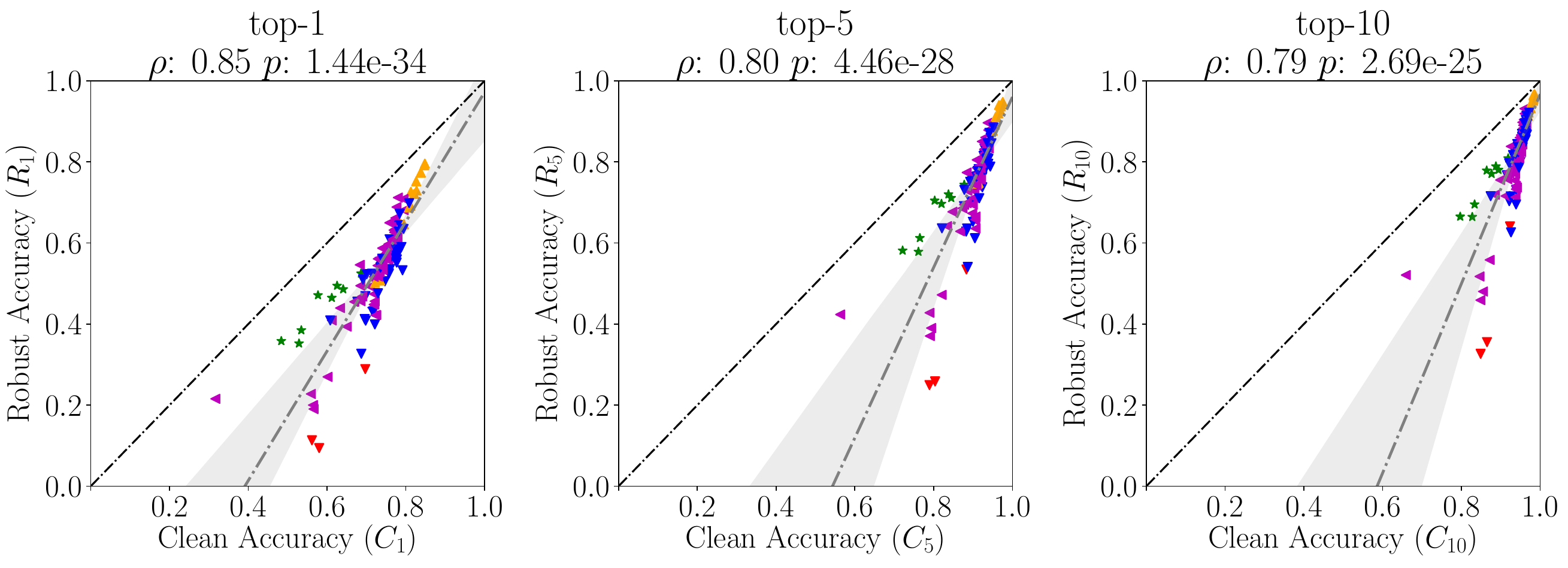}
        \includegraphics[width=0.99\textwidth]{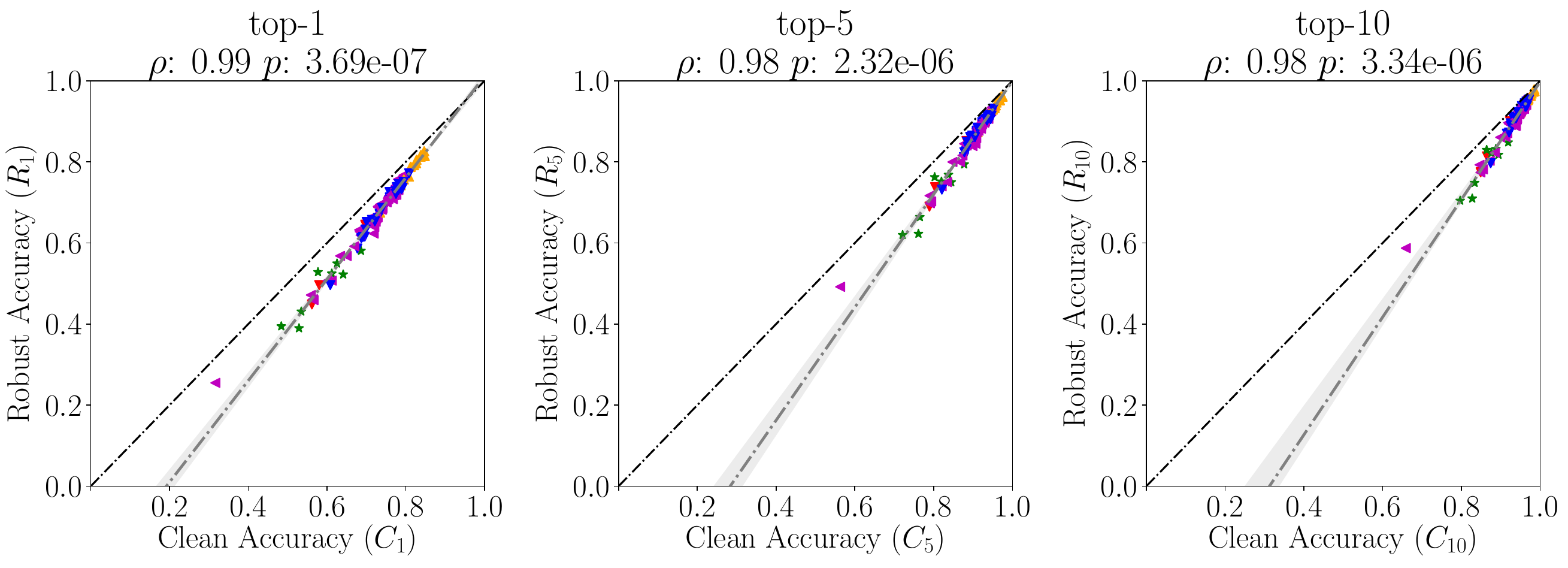}
        \includegraphics[width=0.99\textwidth]{figures/experiments/legend.pdf}
    \caption{Clean vs robust accuracy for adversarial (\textit{top row}) and random (\textit{bottom row}) patches. The Pearson correlation coefficient $\rho$ and the $p$-value are also reported for each plot. The dashed grey line and shaded area show a robust regression model fitted on the data along with the $95\%$ confidence intervals. The results highlight the effectiveness of our pre-optimized strategy against choosing patches at random. }
    \label{fig:results_random_patch_vs_optimized}
\end{figure*}

We now discuss the effectiveness of our dataset on a large-scale setting, where we extend the analysis to a pool of \largescalenummodels models, including also the ones already tested in Section~\ref{subsec:exp_results}.
These are all the models available in RobustBench~\cite{croce2021robustbench} and in ImageNet Testbed~\cite{taori2020measuring}, again divided into the same groups (\standardgroup, \robustgroup, \augmentationgroup and \moredatagroup). We plot our benchmark in Figure~\ref{fig:large-scale}, confirming the results presented in Section~\ref{subsec:exp_results}.
To better highlight the efficacy of our adversarial patches, we also depict the difference in terms of accuracy of these target models scored by applying our pre-optimized patches and randomly-generated ones in Figure~\ref{fig:results_random_patch_vs_optimized}.
The top row shows the results for the pre-optimized patches, while the bottom row focuses on the random ones, and each plot also shows a robust regression line, along with its 95\% confidence interval.

The regression we compute on our metrics highlights meaningful observations we can extract from the benchmark.
First, the robust accuracy of each model evaluated with random patches can be still computed as a linear function of clean accuracy, as shown by the plot of the second row of Figure~\ref{fig:results_random_patch_vs_optimized}.
Hence, the clean accuracy can be seen as an accurate estimator of the robust accuracy when using random patches, similarly to what has been found by Taori \etal~\cite{taori2020measuring}.
However, when we evaluate the robustness with our pre-optimized patches, the relation between robust and clean accuracy slightly diverges from a linear regression model, as the distance of the points from the interpolating line increases.
Such effect is also enforced by the Pearson correlation computed and reported on top of each plot, since it is lower when using adversarial patches.

Among the many reasons behind this effect, we focus on the \robustgroup group, as it lays outside the confidence level, and towards the bisector of the plot, lowering for sure the computed correlation.
Intuitively, models that are located above the regression line can be considered more robust when compared with the others, since their robust accuracy is closer to their clean accuracy, \ie
closer to the bisector line. 
However, even if their robust training is aiding their performances against patch attacks, their robustness is not as evident as the one obtained when considering their original threat model. 
Evaluating adversarial robustness on limited threat models is therefore not sufficient to have a clear idea of what impact attacks can have on these models. Our dataset can help by providing additional analysis of robustness against patch attacks to assess for a more general and complete evaluation.

Lastly, we notice that the \moredatagroup group seems to present a similar effect by distantiating from the regression line, but with a much lower magnitude.
The effect is less evident because these models start from a higher clean accuracy, which then leads to a naturally higher robust accuracy.

\begin{figure*}[t]
    \centering
    \includegraphics[width=0.9\textwidth]{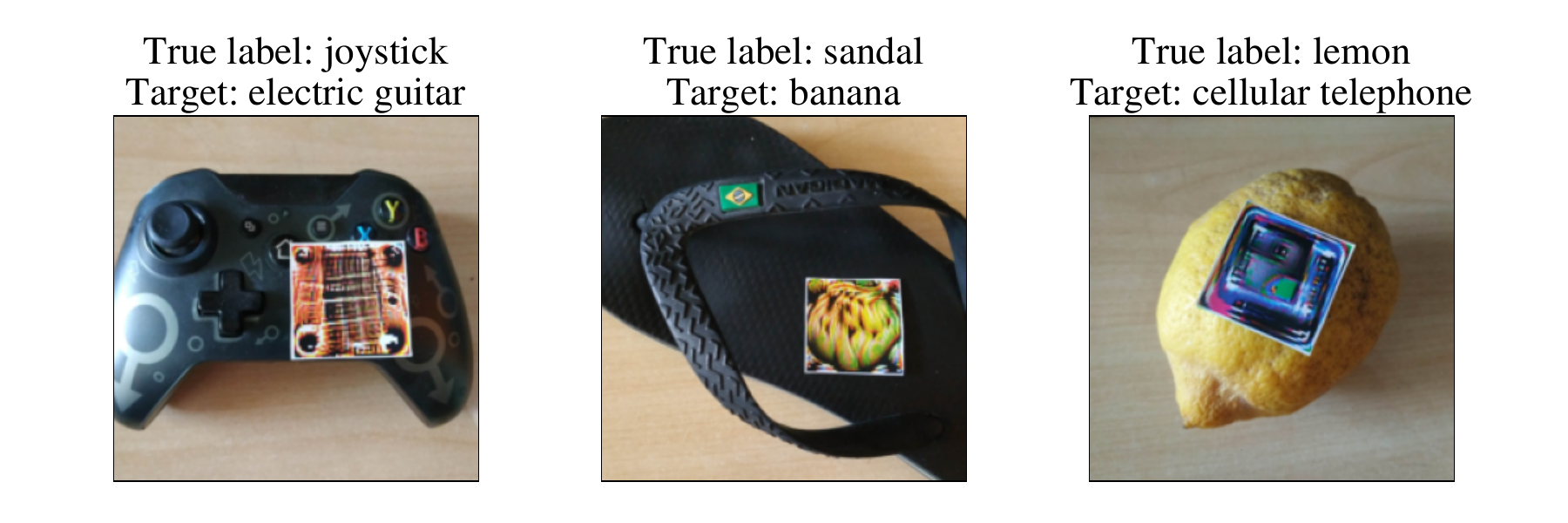}
    \caption{\edit{Examples of adversarial patches from our dataset applied to objects in the physical world. In each photo we show the original and predicted label.}}
    \label{fig:physical_patches}
\end{figure*}

\edit{\subsection{Effectiveness in the physical world}
We now show how our pre-computed patches are effective to assess the robustness of object classification models deployed in the physical world.
To this end, we select 3 objects, i.e., a \emph{joystick}, a \emph{sandal}, and a \emph{lemon}, and we acquire photos of them by applying our 10 patches with 3 different roto-translations, hence composing a dataset of 90 images.
We show some examples of applied patches in Figure~\ref{fig:physical_patches}.
We then select the same models used in Table~\ref{tab:results_groups} (from the  \standardgroup, \robustgroup, \augmentationgroup, and \moredatagroup groups), and report their robust accuracy against such attacks in Table~\ref{tab:physical_results_groups}.}

\begin{table*}[t]
\renewcommand{\arraystretch}{1.5}
\resizebox{0.99\linewidth}{!}{%
\begin{tabular}{l|l|lll|lll|lll|}

\cline{3-11}
\multicolumn{2}{c}{} &  \multicolumn{3}{|c}{\textbf{top-1}} & \multicolumn{3}{c}{\textbf{top-5}} & \multicolumn{3}{c|}{\textbf{top-10}} \\
\cline{2-11}
   &                     \textbf{Model} &            $\cleanacc_{1}$ & $\robustacc_{1}$ & $\successrate_{1}$ & $\cleanacc_{5}$ & $\robustacc_{5}$ & $\successrate_{5}$ & $\cleanacc_{10}$ & $\robustacc_{10}$ & $\successrate_{10}$ \\
\hline \hline
\multirow{3}{*}{\rotatebox[origin=c]{90}{\whiteboxmodelsgroup}}
&                      AlexNet~\cite{krizhevsky2012imagenet} &           0.322 & 0.111 & 0.100 & 0.489 & 0.233 & 0.222 & 0.667 & 0.267 & 0.333 \\
&                     ResNet18~\cite{he2016deep} &           	0.578 & 0.289 & 0.233 & 0.933 & 0.478 & 0.556 & 0.967 & 0.544 & 0.733 \\
&                   SqueezeNet~\cite{iandola2016squeezenet} &          0.456 & 0.222 & 0.344 & 0.744 & 0.322 & 0.589 & 0.944 & 0.422 & 0.722\\
\cline{1-11}
\multirow{3}{*}{\rotatebox[origin=c]{90}{\standardgroup}}
&                    GoogLeNet~\cite{Szegedy2015GoingDW} &           0.422 & 0.311 & 0.067 & 0.767 & 0.378 & 0.367 & 0.933 & 0.456 & 0.489 \\
&                    MobileNet~\cite{Howard2019SearchingFM} &           0.789 & 0.344 & 0.022 & 0.989 & 0.556 & 0.122 & 0.989 & 0.656 & 0.222 \\
&                    Inception v3~\cite{Szegedy2016RethinkingTI} &           0.722 & 0.133 & 0.156 & 0.867 & 0.389 & 0.389 & 0.944 & 0.522 & 0.444 \\
\cline{1-11}
\multirow{3}{*}{\rotatebox[origin=c]{90}{\robustgroup}}
&       Engstrom et al.~\cite{robustness} &           0.333 & 0.222 & 0.044 & 0.722 & 0.411 & 0.156 & 0.922 & 0.522 & 0.178 \\
&            Salman et al.~\cite{DBLP:conf/nips/SalmanIEKM20} &           0.433 & 0.211 & 0.022 & 0.911 & 0.444 & 0.144 & 0.978 & 0.578 & 0.178 \\
&                 Wong et al.~\cite{Wong2020Fast} &           0.311 & 0.078 & 0.033 & 0.678 & 0.267 & 0.133 & 0.733 & 0.422 & 0.167 \\
\cline{1-11}
\multirow{3}{*}{\rotatebox[origin=c]{90}{\texttt{AUGM.}}}
&                Zhang et al.~\cite{zhang2019shiftinvar} &           0.344 & 0.067 & 0.122 & 0.444 & 0.200 & 0.222 & 0.667 & 0.233 & 0.344 \\
&        Hendrycks et al~\cite{hendrycks2021many} &           0.833 & 0.322 & 0.100 & 0.967 & 0.467 & 0.344 & 1.000 & 0.678 & 0.444 \\
& Engstrom et al~\cite{engstrom2019exploring} &           0.722 & 0.289 & 0.133 & 0.933 & 0.422 & 0.322 & 0.956 & 0.511 & 0.467 \\
\cline{1-11}
\multirow{3}{*}{\rotatebox[origin=c]{90}{\moredatagroup}}
&      Yalniz et al.~\cite{yalniz2019billionscale}-a &          0.944 & 0.811 & 0.000 & 1.000 & 0.944 & 0.178 & 1.000 & 0.989 & 0.311 \\
& Yalniz et al.~\cite{yalniz2019billionscale}-b &          1.000 & 1.000 & 0.000 & 1.000 & 1.000 & 0.011 & 1.000 & 1.000 & 0.067 \\
&               Mahajan et al.~\cite{wslimageseccv2018} &          0.733 & 0.389 & 0.067 & 0.922 & 0.678 & 0.289 & 0.967 & 0.800 & 0.356 \\
\bottomrule

\end{tabular}
}
\caption{\edit{Evaluation results of the printed patches applied on the three selected objects (joystick, sandal and lemon).
On the rows, we list the 15 models used for testing, divided into the isolated groups.
On the columns, we detail the clean accuracy, the robust accuracy and the success rate of the adversarial patch, repeated for top-1,5, and 10 accuracy.}}
\label{tab:physical_results_groups}
\end{table*}

\edit{Even if the effectiveness of the printed patches is  lower than their digital counterparts, their efficacy is aligned with the results reported in Table~\ref{tab:results_groups}.
Such performance drop could be caused by the printing quality of the patches, or also by some slight environmental light exposition, that could have altered the colors during the acquisition phase~\cite{eykholt2018robust}.
The \whiteboxmodelsgroup group models are affected more by the application of our patches, as their gradients were used to optimize the attacks, while they show little-to-none efficacy against the \robustgroup group, as expected.
Moreover, both the top-5 and top-10 success rates for the testing groups match the tests conducted in the digital domain, confirming the effectiveness of the given patches also in the physical world.}


\subsection{Discussion}
We briefly summarize here the results of our analysis, based on our \datasetname dataset to benchmark machine-learning models.
We observe that data augmentation techniques do not generally improve robustness to adversarial patches.
Moreover, we argue that real progress in robustness should be observed as a general property against different adversarial attacks, and not only against one specific perturbation model with a given budget (e.g., $\ell_\infty$-norm perturbations with maximum size of $8/255$).
Considering defenses that work against one specific perturbation model may be too myopic and hinder sufficient progress in this area. We are not claiming that work done on defenses for adversarial attacks so far is useless. Conversely, there has been great work and progress in this area, but it seems now that defenses are becoming too specific to current benchmarks and fail to generalize against slightly-different perturbation models. To overcome this issue, we suggest to test the proposed defenses on a wider set of robustness benchmarks, rather than over-specializing them on a specific scenario, and we do believe that our \datasetname benchmark dataset provides a useful contribution in this direction.


\section{Related Work }
\label{sec:sec5_related}
We now discuss relevant work related to the optimization of adversarial patches, and to the proposal of similar benchmark datasets.

\subsection{Patch Attacks}

\begin{table*}[t]
\centering
\resizebox{0.99\textwidth}{!}{
\begin{tabular}{ |c|c|c|c|c|c| } 
 \hline
 Attack & Cross-model & Transfer & Targeted & Untargeted & Transformations \\
 \hline\hline
 Sharif \etal \cite{sharif16-ccs} & \no & \no & \yes & \yes & \rot \\
 Brown \etal \cite{brown2017adversarial} & \yes & \yes & \ding{51} & \no & \loc, \scl, \rot\\
 LaVAN~\cite{karmon2018lavan} & \no & \no & \yes & \no & \loc \\
 PS-GAN~\cite{liu2019perceptual} & \no & \yes & \no & \yes & \loc \\
 DT-Patch~\cite{benz2020double} & \no & \no & \yes & \no & \no \\
 PatchAttack~\cite{yang2020patchattack} & - & \yes & \yes & \yes & \loc, \scl \\
 IAPA~\cite{bai2021inconspicuous} & \no & \yes & \yes & \yes & \no \\
 Lennon \etal~\cite{lennon2021patch} & \no & \yes & \yes & \no & \loc, \scl, \rot \\
 Xiao \etal~\cite{Xiao2021ImprovingTO} & - & \yes & \yes & \yes & \various \\
 Ye \etal~\cite{ye2021patch} & \yes & \yes & \yes & \no & \loc, \scl, \rot \\
 Liu \etal~\cite{liu2020bias} & \no & \yes & \no & \yes & \loc, \scl, \rot \\
 GDPA~\cite{li2021generative} & \no & \no & \yes & \yes & \loc \\
 \edit{Ours (based on \cite{brown2017adversarial})} & \yes & \yes & \yes & \yes & \loc, \rot\\
 \hline
\end{tabular}
}
\caption{Patch attacks, compared based on their main features. \loc refers to the location of the patch in the image, \rot refers to rotation, \scl refers to scale variations, \various include several image transformations (see~\cite{Xiao2021ImprovingTO} for more details).}
\label{tab:patch-attacks}
\end{table*}

The first physical attack against deep neural networks was proposed by~\cite{sharif16-ccs}, by developing an algorithm for printing adversarial eyeglass frames able to evade a face recognition system.
Brown et al.~\cite{brown2017adversarial} introduced the first universal patch attack that focuses on creating a physical perturbation.
Such is obtained by optimizing patches on an ensemble of models to achieve targeted misclassification when applied to different input images with different transformations. 
The LaVAN attack, proposed by Karmon \etal~\cite{karmon2018lavan}, attempts to achieve the same goal of Brown et al. by also reducing the patch size by placing it in regions of the target image where there are no other objects.
The PS-GAN attack, proposed by Liu et al.~\cite{liu2019perceptual}, addresses the problem of minimizing the perceptual sensitivity of the patches by enforcing visual fidelity while achieving the same misclassification objective. 
The DT-Patch attack, proposed by Benz et al.~\cite{benz2020double}, focuses on finding universal patches that only redirect the output of some given classes to different target labels, while retaining normal functioning of the model on the other classes.
PatchAttack, proposed by Yang et al.~\cite{yang2020patchattack}, leverages reinforcement learning for selecting the optimal patch position and texture to use for perturbing the input image for targeted or untargeted misclassification, in a black-box setting. 
The Inconspicuous Adversarial Patch Attack (IAPA), proposed by Bai et al.~\cite{bai2021inconspicuous}, generates difficult-to-detect adversarial patches with one single image by using generators and discriminators.
Lennon et al.~\cite{lennon2021patch} analyze the robustness of adversarial patches and their invariance to 3D poses. 
Xiao et al.~\cite{Xiao2021ImprovingTO} craft transferable patches using a generative model to fool black-box face recognition systems. They use the same transformations as~\cite{Xie2019ImprovingTO}, but unlike other attacks, they apply them to the input image with the patch attached, and not just on the patch.
Ye et al.~\cite{ye2021patch} study the specific application of patch attacks on traffic sign recognition and use an ensemble of models to improve the attack success rate. 
\edit{Liu \etal~\cite{liu2020bias} propose a universal adversarial patch attack that produces patches with strong generalization ability leveraging the texture and semantic bias of the target models to speed up the optimization of the adversarial perturbation. }
The Generative Dynamic Patch Attack (GDPA), proposed by Li et al.~\cite{li2021generative}, generates the patch pattern and location for each input image simultaneously, reducing the runtime of the attack and making it hence a good candidate to use for adversarial training.

We summarize in Table~\ref{tab:patch-attacks} these attacks, highlighting the main properties and comparing them with the attack we used to create the adversarial patches. In particular, in the \emph{Cross-model} column we report the capability of an attack to be performed against multiple models (for black-box attacks we omit this information); in the \emph{Transfer} column the proved transferability of patches, if reported in each work (thus it is still possible that an attack could produce transferable patches even if not tested on this setting); in \emph{Targeted} and \emph{Untargeted} columns the type of misclassification that patches can produce; in \emph{Transformations} column the transformations applied to the patch during the optimization process (if any), which can increase the robustness of the patches with respect to them at test time.

In this work, we leverage the model-ensemble attack proposed by Brown et al.~\cite{brown2017adversarial} to create adversarial patches that are robust to affine transformations and that can be applied to different source images to cause misclassification on different target models. 
From that, we publish a dataset that favors fast robustness evaluation to patch attacks without requiring costly steps for the optimization of the patches, both for the digital and physical world.

\subsection{Benchmarks for Robustness Evaluations}

Previous work proposed datasets for benchmarking adversarial robustness. 
\edit{The APRICOT dataset, proposed by Braunegg et al.~\cite{braunegg2020apricot}, contains $1,000$ annotated photographs of printed adversarial patches targeting object detection systems, \ie producing targeted detections. 
The images are collected in public locations and present different variations in position, distance, lighting conditions, and viewing angle. 
However, even if ImageNet-Patch and APRICOT are similar in spirit, our dataset is designed to test the robustness of image classifiers and not object detectors.
These two problems are very different, and also the techniques used to optimize patches drastically change from one domain to the other.}
ImageNet-C and ImageNet-P, proposed by Hendrycks et al.~\cite{hendrycks2018benchmarking}, are two datasets proposed to benchmark neural network robustness to image corruptions and perturbations, respectively. 
ImageNet-C perturbs images from the ImageNet dataset with a set of 75 algorithmically-generated visual corruptions, including noise, blur, weather, and digital categories, with different strengths. 
ImageNet-P perturbs images again from the ImageNet dataset and contains a sequence of subtle perturbations that slowly perturb the image to assess the stability of the networks' prediction on increasing amounts of perturbations.

Differently from these works, we propose a dataset that can be used to benchmark the robustness of image classifiers to adversarial patch attacks, whose aim is not restricted to being  a source used at training time to improve robustness, or a collection of environmental corruptions. 

\edit{The research community has recently created benchmarks for robustness evaluation of machine-learning models against different attacks.
RobustBench, proposed by Croce \etal~\cite{croce2021robustbench}, provides a standard evaluation protocol for adversarial perturbations and image corruptions. 
The models are then ranked in a leaderboard and downloadable via a dedicated model zoo. 
RobustART, on the other hand, proposed by  Tang\etal~\cite{tang2021robustart}, analyzes the relationship between robustness and different settings including model architectures and training techniques. Our work is the first one to provide a dataset to evaluate the robustness of models against adversarial patch attacks, which can be a nice complement to RobustBench.}

\section{Conclusions, Limitations, and Future Work}
\label{sec:sec6_conclusions}

We propose the \datasetname dataset, a collection of pre-optimized adversarial patches that can be used to compute an approximate-yet-fast robustness evaluation of machine-learning models against patch attacks.
This dataset is constructed by optimizing squared blocks of contiguous pixels perturbed with affine transformations to mislead an ensemble of differentiable models, forcing the optimization algorithm to produce patches that can transfer across models, gaining cross-model effectiveness.
Finally, these adversarial patches are attached to images sampled from the ImageNet dataset, composing a benchmark dataset of 50,000 images.
The latter is used to make an initial robustness evaluation of a selected pool of both standard-trained and robust-trained models, disjointed from the ensemble used to optimize the patch, showing that our methodology is already able to decrease their performances with very few computations needed.
\edit{We also test the effectiveness of our adversarial patches when printed and applied to real-world objects, successfully exhibiting comparable results of their digital counterparts.
Both results highlight the need of considering a wider scope when evaluating adversarial robustness, since the latter should be a general property and not customized on single strategies.
Hence, our dataset can be used to bridge this gap, and to rapidly benchmark the adversarial robustness  of machine-learning models for image classification against patch attacks. }

\myparagraph{Limitations.}
\edit{While our methodology is efficient, it only provides an approximated evaluation of adversarial robustness, which can be computed more accurately by performing adversarial attacks against the target model, instead of using transfer attacks. Hence, our analysis serves as a first preliminary robustness evaluation, to highlight the most promising defensive strategies. Moreover, we only release patches that target 10 different classes, and this number could be extended to target all the 1000 classes of the ImageNet dataset.
%
Lastly, while our methodology only considered the attack proposed by Brown et al.~\cite{brown2017adversarial} to optimize the adversarial patches, it is straightforward to extend our approach and benchmark dataset to also encompass novel and more powerful attacks.}

\myparagraph{Future work.}
\edit{We envision the use of our \datasetname dataset as a benchmark for machine-learning models, which may be  added or used in conjunction with RobustBench. 
We also argue that the proposed methodology is general enough to encompass novel, different patch attacks (e.g., with improved transferability properties~\cite{hang2020ensemble,hu_model_2022}) and image datasets (e.g., MNIST, CIFAR10), thereby easing the creation of novel benchmarks to evaluate robustness against adversarial patches.}

\section{Acknowledgements}
This work was partly supported by the PRIN 2017 project RexLearn, funded by the Italian Ministry of Education, University and Research (grant no. 2017TWNMH2); and by BMK, BMDW, and the Province of Upper Austria in the frame of the COMET Programme managed by FFG in the COMET Module S3AI.




\balance

\end{document}